\documentstyle[12pt]{article}

\newcommand{\be}{\begin{equation}}
\newcommand{\ee}{\end{equation}}
\newcommand{\bea}{\begin{eqnarray}}
\newcommand{\eea}{\end{eqnarray}}
\newcommand{\ep}{\varepsilon}
\newcommand{\Int}{\int\limits}

\newcommand{\nn}{\nonumber}
\newcommand{\half}{{\textstyle{1\over2}}}

\newcommand{\Li}[2]{{\mbox{Li}}_{#1}\left(#2\right)}

\begin{document}
\thispagestyle{empty}
 \begin{flushright}
 INLO--PUB--12/97\\[2mm]
 hep-ph/9712209\\[9mm]
 November 1997
 \end{flushright}
\vspace{30 mm}
\begin{center}
 {\bf \Large
 Threshold \ and \ pseudothreshold \ values  \\[2mm]
 of \ the \ sunset \ diagram}
%  \footnote{This research is supported by the EU under contract
%   number INTAS-93-0744.}
\vspace{10 mm} \\
  F.A.~Berends$^{a,}$
  \footnote{E-mail address: berends@lorentz.leidenuniv.nl}, 
  \ A.I.~Davydychev$^{a,b,}$
  \footnote{E-mail address: davyd@theory.npi.msu.su}
%vsfys1.fi.uib.no}
  \ and \ N.I.~Ussyukina$^{b,}$
  \footnote{E-mail address: ussyuk@theory.npi.msu.su}
\vspace{10 mm} \\
$^{a}${\em 
  Instituut-Lorentz,
  University of Leiden, \\
  P.O.B. 9506, 2300 RA Leiden, The Netherlands}
\vspace{3 mm} \\
$^{b}${\em
  Institute for Nuclear Physics,
  Moscow State University, \\
  119899 Moscow, Russia}
\end{center}
\vspace{12 mm}
\begin{abstract}
Analytic results for the threshold and pseudothreshold values of
the sunset diagram with arbitrary masses are obtained in terms
of dilogarithms of ratios of the masses. 
\end{abstract}

\newpage
\setcounter{page}{2}
\setcounter{footnote}{0}

%========================================================
\section{Introduction}
%========================================================

We shall deal with Feynman integrals corresponding
to the so-called ``sunset'' diagram.
% which is shown in Fig.~1.
This is a two-loop, self-energy-type diagram involving
three propagators. In what follows, we shall need different
integer powers $\nu_i$ of the corresponding denominators. 
The scalar integrals corresponding to this diagram 
with an external momentum $k$ are denoted as
\be
\label{def_L}
L(n; \nu_1,\nu_2,\nu_3) \equiv
\int\int \frac{\mbox{d}^n p \; \mbox{d}^n q}
              {\left[ (p-q)^2 -m_1^2 \right]^{\nu_1}
               \left[ q^2 - m_2^2 \right]^{\nu_2}
               \left[ (k-p)^2-m_3^2 \right]^{\nu_3}} \; ,
\ee
where $n=4-2\ep$ is the space-time dimension in the framework
of dimensional regularization \cite{dimreg}.
The sunset diagram possesses a three-particle threshold
at $k^2=(m_1+m_2+m_3)^2$ and (in general) three pseudothresholds,
at $k^2=(m_1+m_2-m_3)^2$, $(m_1-m_2+m_3)^2$ and $(-m_1+m_2+m_3)^2$.  

There are several reasons why analytical evaluation of such 
diagrams and, in particular, its threshold and peudothreshold 
values, is important: \\
(i) They are needed for calculation of some realistic radiative
corrections in the Standard Model and its extensions 
(in particular, in the Higgs sector). The sunset-type integral
is a part of the basis for any two-loop two-point calculation
(see e.g. in \cite{Tar_new}). \\
(ii) Although this is one of the basic two-loop-order diagrams
in the Quantum Field Theory, analytic results for general
values of the external momentum $k$ and the masses $m_i$ are
not available\footnote{The results for some other two-loop self-energy 
diagrams can be found in refs.~\cite{special,Broadh,BFT,Scharf}.}, 
at least in terms of special functions like
(generalized) polylogarithms. Moreover, there are some arguments
\cite{Scharf} that the results for such diagrams cannot be
expressed in terms of polylogartihms, with a possible exception
of special values of $k^2$. \\
(iii) This is the simplest example of a diagram involving 
a three-particle cut with all the three particles being massive.
Exact results for this diagram would be useful for understanding
the analytic structure of three-particle cuts.
Note that the four-dimensional case is much more complicated than
the three-dimensional one considered in \cite{Rajantie}. \\
(iv) The threshold values of such diagrams are needed for
constructing analytic approximations of the behaviour near the 
threshold, cf. e.g. refs.~\cite{BDS,Tkachov,BS}.\\
(v) In some cases (for example, when two masses are equal, $m_1=m_2$), 
the pseudothreshold may coincide with the on-shell limit
$k^2=m_3^2$ which is relevant for the on-shell calculations 
(see e.g. in refs.~\cite{BG,CzM}). \\      
(vi) Some numerical approaches (in particular, the one described
in \cite{GvdB}) involve the sunset-like integrals as ``kernels''
of integral representations for more general two-loop graphs.
It should be noted that some integral representations which
can be used for numerical calculation of two-loop self-energy
diagrams can also be found in refs.~\cite{numeric,PT}. \\
(vii) The threshold value corresponds to an
infinite sum related to the small momentum expansion
\cite{DT1,FT,Tar}, or the large momentum expansion \cite{DST}, 
taken at its border of convergence. The closed form for 
the coefficients of such expansions (in terms of
generalized hypergeometric functions) was given in ref.~\cite{BBBS}.
Therefore, as a by-product of the calculation one could get
some summation formulae for complicated hypergeometric functions.
Furthermore, since the sunset diagram may be represented in terms of
a one-dimensional integral involving four Bessel-type functions
\cite{Mendels,BBBS}, the analytic results are applicable to 
those integrals, too. 

%=====================================================
\section{Approach to the calculation}
%=====================================================

The standard Feynman parametric representation for the integral
(\ref{def_L}) reads
\bea
\label{Fp1}
L(n; \nu_1, \nu_2, \nu_3)
= \mbox{i}^{2-2n}\; \pi^n \; 
\frac{\Gamma(\nu_1+\nu_2+\nu_3-n)}
     {\Gamma(\nu_1) \; \Gamma(\nu_2) \; \Gamma(\nu_3)}
\Int_0^1 \! \Int_0^1 \! \Int_0^1
\frac{ \prod \alpha_i^{\nu_i-1} \mbox{d}\alpha_i \; \;
       \delta\left(\sum\alpha_i-1\right)}
     {(\alpha_1 \alpha_2 + \alpha_1 \alpha_3 
                         + \alpha_2 \alpha_3)^{3n/2-\Sigma\nu_i}}
\nn \\
\times 
\frac{1}{\left[\alpha_1 \alpha_2 \alpha_3 k^2 - 
         (\alpha_1 \alpha_2 + \alpha_1 \alpha_3 + \alpha_2 \alpha_3)
         (\alpha_1 m_1^2 + \alpha_2 m_2^2 + \alpha_3 m_3^2)
         \right]^{\Sigma\nu_i-n}} .
\eea
Using tricks similar to those described in \cite{Scharf},
we can get rid of the first denominator in the integrand of
eq.~(\ref{Fp1}). Namely, let us use {\em exactly} the same 
transformation of $\alpha$-variables  (first inverting and then 
rescaling) as in eq.~(13) of \cite{DT2},
\be
\alpha_i=(\alpha'_i)^{-1}, \hspace{5mm}
\alpha'_i={\cal{F}}({\alpha''}_1,{\alpha''}_2,{\alpha''}_3),
\hspace{5mm}
{\cal{F}}({\alpha''}_1,{\alpha''}_2,{\alpha''}_3)
= \frac{{\alpha''}_1^{-1} + {\alpha''}_2^{-1} + {\alpha''}_3^{-1}}
       {{\alpha''}_1 + {\alpha''}_2 + {\alpha''}_3} .
\ee
Suppressing the primes, we arrive at the following modified 
representation:
\bea
\label{Fp2}
L(n; \nu_1, \nu_2, \nu_3)
= \mbox{i}^{2-2n}\; \pi^n \; 
\frac{\Gamma(\nu_1+\nu_2+\nu_3-n)}
     {\Gamma(\nu_1) \; \Gamma(\nu_2) \; \Gamma(\nu_3)}
\hspace{60mm}
\nn \\
\times
\Int_0^1 \! \Int_0^1 \! \Int_0^1
\frac{ \alpha_1^{\nu_2+\nu_3-n/2-1} \alpha_2^{\nu_1+\nu_3-n/2-1}
           \alpha_3^{\nu_1+\nu_2-n/2-1} \;
       \prod \mbox{d}\alpha_i \; \;
       \delta\left(\sum\alpha_i-1\right)}
     {\left[\alpha_1 \alpha_2 \alpha_3 k^2 - 
         \alpha_2 \alpha_3 m_1^2 - \alpha_1 \alpha_3 m_2^2 
          - \alpha_1 \alpha_2 m_3^2
         \right]^{\Sigma\nu_i-n}} .
\eea
The equivalence of the representations (\ref{Fp1}) and (\ref{Fp2})
can also be established by comparing the corresponding triple
Mellin--Barnes contour integrals. 

Furthermore, using Cheng--Wu theorem \cite{ChengWu} (see
also in \cite{BFO2}, Appendix~B) 
and rescaling the variables,
one can transform the representation
(\ref{Fp2}) into
\bea
\label{CW2}
L(n; \nu_1, \nu_2, \nu_3)
= \mbox{i}^{2-2\Sigma\nu_i}\; \pi^n \; 
\frac{\Gamma(\nu_1+\nu_2+\nu_3-n)}
     {\Gamma(\nu_1) \; \Gamma(\nu_2) \; \Gamma(\nu_3)} \;
\left( \prod m_i^{n/2-\nu_i} \right)
\hspace{40mm}
\nn \\
\times
\Int_0^{\infty}\! \Int_0^{\infty}
\frac{\mbox{d}\xi \; \mbox{d}\eta \;\; 
       \xi^{\nu_2+\nu_3-n/2-1} \; \eta^{\nu_1+\nu_3-n/2-1}}
     {(m_1\xi\!+\!m_2\eta\!+\!m_3)^{3n/2-\Sigma\nu_i}
\left[ 
(m_1\xi\!+\!m_2\eta\!+\!m_3)(m_1\eta\!+\!m_2\xi\!+\!m_3\xi\eta)
             -k^2 \xi\eta \right]^{\Sigma\nu_i-n}} .
\eea

The cubic form in the denominator has the following representations
at the pseudothresholds and at the threshold:
\bea
\label{cubic}
\left[ (m_1\xi+m_2\eta+m_3)(m_1\eta+m_2\xi+m_3\xi\eta)-k^2 \xi\eta 
\right]
\hspace{63mm}
\nn \\
= m_2 m_3 \xi (1-\eta)^2 
+ m_1 m_3 \eta (1-\xi)^2 + m_1 m_2 (\xi-\eta)^2
  + \left( (m_1+m_2+m_3)^2 - k^2 \right) \xi \eta
\hspace{4mm}
\nn \\
= \left\{
\begin{array}{l}
m_2 m_3 \xi (1-\eta)^2 + m_1 m_3 \eta (1+\xi)^2 
+ m_1 m_2 (\xi+\eta)^2,
\hspace{5mm} k^2=(-m_1+m_2+m_3)^2 \\
m_2 m_3 \xi (1+\eta)^2 
+ m_1 m_3 \eta (1-\xi)^2 + m_1 m_2 (\xi+\eta)^2,
\hspace{5mm} k^2=(m_1-m_2+m_3)^2 \\
m_2 m_3 \xi (1+\eta)^2 
+ m_1 m_3 \eta (1+\xi)^2 + m_1 m_2 (\xi-\eta)^2,
\hspace{5mm} k^2=(m_1+m_2-m_3)^2 \\
m_2 m_3 \xi (1-\eta)^2 
+ m_1 m_3 \eta (1-\xi)^2 + m_1 m_2 (\xi-\eta)^2,
\hspace{5mm} k^2=(m_1+m_2+m_3)^2 
\end{array}
\right.
\eea
In particular, it can be seen that this cubic form is positive
semidefinite at the threshold and pseudothresholds.

Using eq.~(\ref{Fp2}), it is easy to derive
the following decomposition:
\bea
\label{decomp111}
L(4-2\ep; 1,1,1) = \frac{1}{1-2\ep} 
\left\{ 
%k^2 \frac{\partial}{\partial k^2} L(4-2\ep; 1,1,1)
-k^2 \pi^{-2} L(6-2\ep; 2,2,2)
\hspace{59mm}
\right.
\nn \\
\left.
+ m_1^2 L(4-2\ep; 2,1,1) + m_2^2 L(4-2\ep; 1,2,1) 
+ m_3^2 L(4-2\ep; 1,1,2)
\frac{}{} \right\} .
\eea
This decomposition is similar to one used in \cite{GvdB}. 
Taking into account that
\be
L(6-2\ep; 2,2,2)
=-\pi^2 \frac{\partial}{\partial k^2} L(4-2\ep; 1,1,1) ,
\ee
eq.~(\ref{decomp111}) just reflects the fact that the 
mass-squared dimension
of the integral $L(4-2\ep;1,1,1)$ is $(1-2\ep)$.

The calculation of each of the integrals on the r.h.s. 
of (\ref{decomp111})
is simpler than direct calculation of $L(4-2\ep; 1,1,1)$.
Using the representation (\ref{CW2}), we see that the integrals
$L(6-2\ep; 2,2,2)$ and $L(4-2\ep; 1,1,2)$ are proportional to
\be
\label{r}
\Gamma(2\ep) \Int_0^{\infty} \Int_0^{\infty}
\frac{\mbox{d}\xi\; \mbox{d}\eta\; \; \xi^{\ep} \; \eta^{\ep}}
     {(m_1\xi + m_2\eta + m_3)^{r-3\ep} \; 
           \left[ \mbox{cubic form} \right]^{2\ep}} ,
\ee 
where $r=3$ for $L(6-2\ep;2,2,2)$ and $r=2$ for $L(4-2\ep; 1,1,2)$.

When $r=3$, the double integral in (\ref{r})  
is convergent, and what we need is just to expand the integrand
in $\ep$, keeping the terms of order $\ep$ (since we have got
a singular factor $\Gamma(2\ep)$ in front of the integral).
In this way, we get
\bea
\label{r3}
\Gamma(2\ep) \Int_0^{\infty} \Int_0^{\infty}
\frac{\mbox{d}\xi\; \mbox{d}\eta}
     {(m_1\xi + m_2\eta + m_3)^3} \; 
\left\{ 1+ \ep \ln\xi +\ep \ln\eta +3\ep \ln(m_1\xi +m_2\eta +m_3)
\right.
\nn  \\
\left.
- 2\ep \ln[\mbox{cubic form}] \right\} + {\cal{O}}(\ep).
\eea
 
When $r=2$, the integral (\ref{r}) develops a ($1/\ep$) singularity
as $\xi,\eta\to\infty$. One can
subtract from (\ref{r}) a simpler integral with the same
asymptotic behaviour as $\xi,\eta\to\infty$, 
\be
\label{r_sub}
\Gamma(2\ep) \Int_0^{\infty} \Int_0^{\infty}
\frac{\mbox{d}\xi\; \mbox{d}\eta\; \; \xi^{\ep} \; \eta^{\ep}}
     {(m_1\xi + m_2\eta + m_3)^{r-3\ep} \; 
           \left[ (m_1\xi + m_2\eta) m_3 \xi\eta \right]^{2\ep}} ,
\ee 
which can be calculated in terms of $\Gamma$ functions.
The difference of (\ref{r}) and (\ref{r_sub}) is convergent
and yields
\be
\label{r2}
\Int_0^{\infty} \Int_0^{\infty}
\frac{\mbox{d}\xi\; \mbox{d}\eta}
     {(m_1\xi + m_2\eta + m_3)^2} \; 
\ln\left( 
\frac{(m_1\xi + m_2\eta) m_3 \xi\eta}
     {\mbox{[cubic form]}} \right) + {\cal{O}}(\ep) .
\ee

When we consider representation (\ref{CW2}) for  
$L(4-2\ep;2,1,1)$ and $L(4-2\ep;1,2,1)$, it does not diverge as
$\xi,\eta\to\infty$. Instead, it does diverge as $\xi\to~0$ or
$\eta\to~0$. However, these integrals can be reduced to
$L(4-2\ep;1,1,2)$ just by a permutation of $m_1, m_2, m_3$
and thus calculated via subtractions similar to (\ref{r_sub}).

The most complicated point in calculating the parametric integrals
(\ref{r3}) and (\ref{r2}) is how to deal with the contributions
involving $\ln[\mbox{cubic~form}]$. One of possible ways
is to substitute the variables,
\be
\xi=\frac{\sigma}{m_1+m_2}(m_3+m_2 \rho), \; \;
\eta=\frac{\sigma}{m_1+m_2}(m_3-m_1 \rho) ,
\ee
and then integrate over $\rho$ between $(-m_3/m_2)$ and 
$(m_3/m_1)$. Depending on a region (and whether a pseudothreshold
or the threshold is considered), this integral yields an arctangent 
or a hyperbolic arctangent (the latter can be presented as a 
logarithm) 
of an argument involving square roots. The most labour-consuming
part was to calculate the remaining integral over $\sigma$,
which was done (for the cases considered below) by using
some tricky trigonometric substitutions.  

%=========================================================
\section{General results at the pseudothreshold}
%=========================================================

We shall consider the pseudothreshold at $k^2=(m_1+m_2-m_3)^2$.
The results for the other two pseudothresholds,
at $k^2=(m_1-m_2+m_3)^2$ and $k^2=(-m_1+m_2+m_3)^2$ 
can be obtained by permutation of the indices 1,2,3.

It is convenient to introduce the following dilogarithmic 
functions\footnote{Up to $\ln^2{z}$ terms, these functions
are similar to those used in refs.~\cite{Broadh,BG,CzM}.}:
\bea
\label{T-}
T^{-}(z) &=& \Li{2}{-z} - \Li{2}{-1/z}
            +\ln z \ln\left( (1+z)^2/z \right)
\nn \\
&=& 2 \Li{2}{-z} + {\textstyle{1\over6}}\pi^2 + 2\ln z \ln(1+z) 
    - \half \ln^2 z
\nn \\
&=& 2 \Li{2}{ 1/(1+z) } -  {\textstyle{1\over6}}\pi^2
    + \ln^2(1+z) - \half \ln^2 z ,
\eea
\bea
\label{T+}
T^{+}(z) &=& \Li{2}{1-1/z} - \Li{2}{1-z}
\nn \\
&=& 2 \Li{2}{z} -  {\textstyle{1\over3}}\pi^2 + 2\ln z \ln(1-z) 
    - \half \ln^2 z
\nn \\
&=& 2 \Li{2}{ 1/(1-z) } + \half \ln^2 z ,
\eea
which are antisymmetric under inversion,
$T^{\pm}(1/z) = - T^{\pm}(z)$.
In particular, $T^{\pm}(1)=0$ and 
$T^{\pm}(m_j/m_l)=-T^{\pm}(m_l/m_j)$.
Note that the functions $T^{\pm}(z)$ continue each other
to the region of negative $z$'s, namely
$T^{\pm}(z)\leftrightarrow T^{\mp}(-z)$.

Let us list the results for the integrals involved on the r.h.s. of 
eq.~(\ref{decomp111}):
\bea
\label{L6222p}
\left. \frac{}{} L(6-2\ep;2,2,2)\right|_{k^2=(m_1+m_2-m_3)^2} 
= \pi^{6-2\ep} \Gamma^2(1+\ep) 
\hspace{63mm}
\nn \\
\times\!\left\{
-\frac{1}{4 \ep} \;
\frac{m_1^{1-4\ep}+m_2^{1-4\ep}-m_3^{1-4\ep}}{m_1+m_2-m_3}
-\frac{9}{8}+\frac{m_1^2+m_2^2+m_3^2}{2 (m_1+m_2-m_3)^2}
\right.
\hspace{46mm}
\nn \\
+\frac{m_1 m_2 m_3}{(m_1\!+\!m_2\!-\!m_3)^3}
\left[ \frac{m_1\!-\!m_2}{m_3} \; \ln\frac{m_1}{m_2}
      -\frac{m_2\!+\!m_3}{m_1} \; \ln\frac{m_2}{m_3} 
      -\frac{m_1\!+\!m_3}{m_2} \; \ln\frac{m_1}{m_3}
\right]
\hspace{22mm}
\nn \\
+ \frac{m_1^2 m_2^2 m_3^2}{(m_1\!+\!m_2\!-\!m_3)^4}
\left[
 \frac{m_1^2\!-\!m_2^2}{m_1^2 m_2^2}\; 
T^{-}\left(\frac{m_1}{m_2}\right)
+\frac{m_2^2\!-\!m_3^2}{m_2^2 m_3^2}\; 
T^{+}\left(\frac{m_2}{m_3}\right)
+\frac{m_3^2\!-\!m_1^2}{m_3^2 m_1^2}\; 
T^{+}\left(\frac{m_3}{m_1}\right)
\right.
\hspace{3mm}
\nn \\
+ \left( \frac{1}{m_3^2} - \frac{1}{2m_1^2} -\frac{1}{2m_2^2} \right) 
\left( \ln^2\frac{m_1}{m_2}+ \frac{\pi^2}{3} \right)
\hspace{52mm}
\nn \\
\left. \left. 
+ \left( 
\frac{1}{m_1^2}\!-\!\frac{1}{2m_2^2}\!-\!\frac{1}{2m_3^2} \right) 
         \ln^2\frac{m_2}{m_3} 
+ \left( 
\frac{1}{m_2^2}\!-\!\frac{1}{2m_3^2}\!-\!\frac{1}{2m_1^2} \right) 
         \ln^2\frac{m_3}{m_1} \right]
\right\}\!+\!{\cal{O}}(\ep),
\eea
\bea
\label{L4112p}
\left. \frac{}{} L(4-2\ep;1,1,2)\right|_{k^2=(m_1+m_2-m_3)^2} 
= \pi^{4-2\ep} \; \frac{\Gamma^2(1+\ep)}{1-\ep} \; m_3^{-4\ep}
\hspace{51mm}
\nn \\
\times\!\left\{
-\frac{1}{2\ep^2}+1+\frac{\pi^2}{3}
+T^{+}\left(\frac{m_1}{m_3}\right)
+T^{+}\left(\frac{m_2}{m_3}\right)
+\frac{1}{2}\ln^2\frac{m_1}{m_3}
+\frac{1}{2}\ln^2\frac{m_2}{m_3} 
\right.
\hspace{32mm}
\nn \\
-\frac{2}{m_1\!+\!m_2\!-\!m_3}
\left[ m_1 \ln\frac{m_1}{m_3}+m_2 \ln\frac{m_2}{m_3} \right]
\hspace{81mm}
\nn \\
+\frac{1}{(m_1\!+\!m_2\!-\!m_3)^2}
\left[ 
(m_1^2-m_2^2) T^{-}\left(\frac{m_1}{m_2}\right)
- m_1^2 T^{+}\left(\frac{m_2}{m_3}\right)
- m_2^2 T^{+}\left(\frac{m_1}{m_3}\right)
\right.
\hspace{24mm}
\nn \\
\left. \left.
-\frac{m_1^2\!+\!m_2^2}{2} 
\left( \ln^2\frac{m_1}{m_2} \!+\!\frac{\pi^2}{3} \right)
+\frac{2 m_2^2\!-\!m_1^2}{2} \ln^2\frac{m_2}{m_3} 
+\frac{2 m_1^2\!-\!m_2^2}{2} \ln^2\frac{m_1}{m_3}
\right] \right\} + {\cal{O}}(\ep),
\eea
\bea
\label{L4211p}
\left. \frac{}{} L(4-2\ep;2,1,1)\right|_{k^2=(m_1+m_2-m_3)^2} 
=\pi^{4-2\ep} \frac{\Gamma^2(1+\ep)}{1-\ep} \; m_1^{-4\ep}
\hspace{51mm}
\nn \\
\times \! \left\{
-\frac{1}{2\ep^2}+1
- \frac{\pi^2}{6}
- T^{-}\left(\frac{m_1}{m_2}\right)
- T^{+}\left(\frac{m_1}{m_3}\right)
+ \frac{1}{2} \ln^2\frac{m_1}{m_2}
+ \frac{1}{2} \ln^2\frac{m_1}{m_3}
\right.
\hspace{30mm}
\nn \\
+ \frac{2}{m_1\!+\!m_2\!-\!m_3} 
\left[ m_2 \ln\frac{m_1}{m_2}+m_3 \ln\frac{m_3}{m_1} \right]
\hspace{79mm}
\nn \\
+\frac{1}{(m_1\!+\!m_2\!-\!m_3)^2}
\left[ 
m_3^2 T^{-}\left(\frac{m_1}{m_2}\right)
+(m_2^2\!-\!m_3^2) T^{+}\left(\frac{m_2}{m_3}\right)
+ m_2^2 T^{+}\left(\frac{m_1}{m_3}\right)
\right.
\hspace{22mm}
\nn \\
\left. \left.   
+ \frac{2 m_2^2\!-\!m_3^2}{2} 
      \left( \ln^2\frac{m_1}{m_2}\!+\!\frac{\pi^2}{3} \right)
- \frac{m_2^2\!+\!m_3^2}{2} \ln^2\frac{m_2}{m_3} 
+ \frac{2 m_3^2\!-\!m_2^2}{2} \ln^2\frac{m_1}{m_3}
\right] \right\} \!+\! {\cal{O}}(\ep).
\eea
The result for $L(4-2\ep;1,2,1)$ can be obtained from (\ref{L4211p})
by permutation $m_1\leftrightarrow m_2$. Substituting 
eqs.~(\ref{L6222p})--(\ref{L4211p}) into eq.~(\ref{decomp111}) 
we get
\bea
\label{L4111p}
\left. \frac{}{} L(4-2\ep;1,1,1)\right|_{k^2=(m_1+m_2-m_3)^2} 
= \pi^{4-2\ep} \frac{\Gamma^2(1+\ep)}{(1-\ep)(1-2\ep)}
\hspace{51mm}
\nn \\
\times\! \left\{
-\frac{1}{2\ep^2} \; 
\left(m_1^{2-4\ep}+m_2^{2-4\ep}+m_3^{2-4\ep}\right)
+\frac{1}{4\ep} (m_1+m_2-m_3) 
\left( m_1^{1-4\ep}+m_2^{1-4\ep}-m_3^{1-4\ep} \right)
\right.
\hspace{1mm}
\nn \\
  +\frac{7}{8} (m_1+m_2-m_3)^2
  +\frac{1}{2} (m_1^2+m_2^2+m_3^2)
  -\frac{\pi^2}{6}(m_1^2+m_2^2-2 m_3^2) 
\hspace{28mm}
\nn \\
-(m_1^2-m_2^2) T^{-}\left(\frac{m_1}{m_2}\right)
-(m_2^2-m_3^2) T^{+}\left(\frac{m_2}{m_3}\right)
-(m_3^2-m_1^2) T^{+}\left(\frac{m_3}{m_1}\right)
\hspace{20mm}
\nn \\
  +\frac{m_1^2+m_2^2}{2} \; \ln^2\frac{m_1}{m_2}
  +\frac{m_2^2+m_3^2}{2} \; \ln^2\frac{m_2}{m_3}
  +\frac{m_3^2+m_1^2}{2} \; \ln^2\frac{m_3}{m_1}
\hspace{38mm}
\nn \\ 
+\frac{m_1 m_2 m_3}{m_1\!+\!m_2\!-\!m_3}
\left[ \frac{m_1-m_2}{m_3} \ln\frac{m_1}{m_2}
      -\frac{m_2+m_3}{m_1} \ln\frac{m_2}{m_3} 
      -\frac{m_1+m_3}{m_2} \ln\frac{m_1}{m_3}
\right]
\hspace{24mm}
\nn \\
+\frac{m_1^2 m_2^2 m_3^2}{(m_1\!+\!m_2\!-\!m_3)^2}
\left[
 \frac{m_1^2\!-\!m_2^2}{m_1^2 m_2^2}\; 
T^{-}\left(\frac{m_1}{m_2}\right)
+\frac{m_2^2\!-\!m_3^2}{m_2^2 m_3^2}\; 
T^{+}\left(\frac{m_2}{m_3}\right)
+\frac{m_3^2\!-\!m_1^2}{m_1^2 m_3^2}\; 
T^{+}\left(\frac{m_3}{m_1}\right)
\right.
\hspace{3mm}
\nn \\
+ \left( \frac{1}{m_3^2} - \frac{1}{2m_1^2} 
-\frac{1}{2m_2^2} \right) 
\left( \ln^2\frac{m_1}{m_2}+ \frac{\pi^2}{3} \right)
\hspace{52mm}
\nn \\
\left. \left. 
+ \left( \frac{1}{m_1^2}\!-\!\frac{1}{2m_2^2}
\!-\!\frac{1}{2m_3^2} \right) 
         \ln^2\frac{m_2}{m_3} 
+ \left( \frac{1}{m_2^2}\!-\!\frac{1}{2m_3^2}
\!-\!\frac{1}{2m_1^2} \right) 
         \ln^2\frac{m_1}{m_3} \right]
\right\} \!+\! {\cal{O}}(\ep). 
\eea

%===================================================
\section{General results at the threshold}
%===================================================

In addition to the notation (\ref{T-}), we shall also introduce
the angles $\theta_1$, $\theta_2$ and $\theta_3$ via
\be
\label{theta_i}
\theta_i \equiv
\arctan\left( m_i \sqrt{\frac{m_1+m_2+m_3}{m_1 m_2 m_3}} \right) .
\ee   
Note that $\theta_1+\theta_2+\theta_3=\pi$. Therefore, they
can be undestood as the angles of a certain triangle 
(cf. e.g. in \cite{DD}).

The results for the threshold values of the integrals
occurring on the r.h.s. of eq.~(\ref{decomp111}) are
\bea
\label{L6222t}
\left. \frac{}{} L(6-2\ep;2,2,2)\right|_{k^2=(m_1+m_2+m_3)^2} 
= \pi^{6-2\ep} \Gamma^2(1+\ep) 
\hspace{63mm}
\nn \\
\times\!\left\{
-\frac{1}{4 \ep} \;
\frac{m_1^{1-4\ep}+m_2^{1-4\ep}+m_3^{1-4\ep}}{m_1+m_2+m_3}
-\frac{9}{8}+\frac{m_1^2+m_2^2+m_3^2}{2 (m_1+m_2+m_3)^2}
\right.
\hspace{46mm}
\nn \\
+\frac{m_1 m_2 m_3}{(m_1\!+\!m_2\!+\!m_3)^3}
\left[ \frac{m_1\!-\!m_2}{m_3} \; \ln\frac{m_1}{m_2}
      +\frac{m_2\!-\!m_3}{m_1} \; \ln\frac{m_2}{m_3}
      +\frac{m_3\!-\!m_1}{m_2} \; \ln\frac{m_3}{m_1} \right]
\hspace{20mm}
\nn \\
+ \frac{m_1^2 m_2^2 m_3^2}{(m_1\!+\!m_2\!+\!m_3)^4}
\left[
 \frac{m_1^2\!-\!m_2^2}{m_1^2 m_2^2}\; 
T^{-}\left(\frac{m_1}{m_2}\right)
+\frac{m_2^2\!-\!m_3^2}{m_2^2 m_3^2}\; 
T^{-}\left(\frac{m_2}{m_3}\right)
+\frac{m_3^2\!-\!m_1^2}{m_3^2 m_1^2}\; 
T^{-}\left(\frac{m_3}{m_1}\right)
\right.
\hspace{3mm}
\nn \\
+ \left( \frac{1}{m_3^2} - \frac{1}{2m_1^2} -\frac{1}{2m_2^2} \right) 
         \ln^2\frac{m_1}{m_2}
+ \left( \frac{1}{m_1^2}\!-\!\frac{1}{2m_2^2}
         \!-\!\frac{1}{2m_3^2} \right) 
         \ln^2\frac{m_2}{m_3} 
\hspace{15mm}
\nn \\
+ \left( \frac{1}{m_2^2}\!-\!\frac{1}{2m_3^2}
\!-\!\frac{1}{2m_1^2} \right) 
         \ln^2\frac{m_3}{m_1}
+\frac{4\pi^2}{3} 
\left( \frac{1}{m_1^2}+\frac{1}{m_2^2}+\frac{1}{m_3^2}\right)
\hspace{24mm}
\nn \\
\left. \left. 
-4 \pi \left( \frac{\theta_1}{m_1^2}\!+\! \frac{\theta_2}{m_2^2}
              \!+\! \frac{\theta_3}{m_3^2} \right)
+4 \pi \left(\frac{1}{m_1}\!+\!\frac{1}{m_2}\!+\!\frac{1}{m_3}\right) 
       \sqrt{\frac{m_1\!+\!m_2\!+\!m_3}{m_1 m_2 m_3}} 
\right]
\right\}\!+\!{\cal{O}}(\ep),
\eea
\bea
\label{L4112t}
\left. \frac{}{} L(4-2\ep;1,1,2)\right|_{k^2=(m_1+m_2+m_3)^2} 
= \pi^{4-2\ep} \; \frac{\Gamma^2(1+\ep)}{1-\ep} \; m_3^{-4\ep}
\hspace{51mm}
\nn \\
\times\!\left\{
-\frac{1}{2\ep^2}+1
+\frac{4\pi^2}{3} - 4\pi\theta_3
+ T^{-}\left(\frac{m_1}{m_3}\right)
+ T^{-}\left(\frac{m_2}{m_3}\right)
+ \frac{1}{2}\ln^2\frac{m_1}{m_3}
+\frac{1}{2} \ln^2\frac{m_2}{m_3} 
\right.
\hspace{16mm}
\nn \\
-\frac{2}{m_1\!+\!m_2\!+\!m_3}
\left[ m_1 \ln\frac{m_1}{m_3}+m_2 \ln\frac{m_2}{m_3} \right]
\hspace{80mm}
\nn \\
+\frac{1}{(m_1\!+\!m_2\!+\!m_3)^2}
\left[ 
(m_1^2-m_2^2) T^{-}\left(\frac{m_1}{m_2}\right)
-m_1^2 T^{-}\left(\frac{m_2}{m_3}\right)
-m_2^2 T^{-}\left(\frac{m_1}{m_3}\right)
\right.
\hspace{23mm}
\nn \\
-\frac{m_1^2+m_2^2}{2} \ln^2\frac{m_1}{m_2} 
+\frac{2 m_1^2-m_2^2}{2} \ln^2\frac{m_1}{m_3}
+\frac{2 m_2^2-m_1^2}{2} \ln^2\frac{m_2}{m_3}
\hspace{20mm}
\nn \\
\left. \left.
+\frac{4\pi^2}{3} \left( m_1^2\!+\!m_2^2 \right)
\!-\!4\pi\left( m_2^2 \theta_1\!+\! m_1^2 \theta_2 \right)
\!-\!4 \pi \sqrt{m_1 m_2 m_3 (m_1\!+\!m_2\!+\!m_3)} 
\right] \right\} \!+\! {\cal{O}}(\ep).
\eea
The results for $L(4-2\ep;2,1,1)$ and $L(4-2\ep;1,2,1)$
can be obtained from (\ref{L4112t}) by permutations 
$m_3\leftrightarrow m_1$
and $m_3\leftrightarrow m_2$, respectively.
Using eq.~(\ref{decomp111}), we get
\bea
\label{L4111t}
\left. \frac{}{} L(4-2\ep;1,1,1)\right|_{k^2=(m_1+m_2+m_3)^2} 
= \pi^{4-2\ep} \frac{\Gamma^2(1+\ep)}{(1-\ep)(1-2\ep)}
\hspace{52.5mm}
\nn \\
\times\! \left\{
-\frac{1}{2\ep^2} \; 
\left(m_1^{2-4\ep}+m_2^{2-4\ep}+m_3^{2-4\ep}\right)
+\frac{1}{4\ep} (m_1+m_2+m_3) 
\left( m_1^{1-4\ep}+m_2^{1-4\ep}+m_3^{1-4\ep} \right)
\right.
\hspace{1.5mm}
\nn \\
  +\frac{7}{8} (m_1+m_2+m_3)^2
  + \left( \frac{1}{2} 
               + \frac{4\pi^2}{3} \right) (m_1^2+m_2^2+m_3^2)
- 4\pi\left(m_1^2 \theta_1 + m_2^2 \theta_2 + m_3^2 \theta_3 \right) 
\hspace{5mm}
\nn \\
-(m_1^2-m_2^2) T^{-}\left(\frac{m_1}{m_2}\right)
-(m_2^2-m_3^2) T^{-}\left(\frac{m_2}{m_3}\right)
-(m_3^2-m_1^2) T^{-}\left(\frac{m_3}{m_1}\right)
\hspace{27mm}
\nn \\
  +\frac{m_1^2\!+\!m_2^2}{2} \ln^2\frac{m_1}{m_2}
  \!+\!\frac{m_2^2\!+\!m_3^2}{2} \ln^2\frac{m_2}{m_3}
  \!+\!\frac{m_3^2\!+\!m_1^2}{2} \ln^2\frac{m_3}{m_1}
  \!-\!4\pi\sqrt{m_1 m_2 m_3 (m_1\!+\!m_2\!+\!m_3)}
\hspace{1mm}
\nn \\ 
+\frac{m_1 m_2 m_3}{m_1\!+\!m_2\!+\!m_3}
\left[ \frac{m_1-m_2}{m_3} \ln\frac{m_1}{m_2}
      +\frac{m_2-m_3}{m_1} \ln\frac{m_2}{m_3} 
      +\frac{m_3-m_1}{m_2} \ln\frac{m_3}{m_1}
\right]
\hspace{25.5mm}
\nn \\
+\frac{m_1^2 m_2^2 m_3^2}{(m_1\!+\!m_2\!+\!m_3)^2}
\left[
 \frac{m_1^2\!-\!m_2^2}{m_1^2 m_2^2}\; 
T^{-}\left(\frac{m_1}{m_2}\right)
+\frac{m_2^2\!-\!m_3^2}{m_2^2 m_3^2}\; 
T^{-}\left(\frac{m_2}{m_3}\right)
+\frac{m_3^2\!-\!m_1^2}{m_1^2 m_3^2}\; 
T^{-}\left(\frac{m_3}{m_1}\right)
\right.
\hspace{3mm}
\nn \\
+ \left( \frac{1}{m_3^2} - \frac{1}{2m_1^2} 
                                  -\frac{1}{2m_2^2} \right) 
         \ln^2\frac{m_1}{m_2}
+ \left( \frac{1}{m_1^2}\!-\!\frac{1}{2m_2^2}
              \!-\!\frac{1}{2m_3^2} \right) 
         \ln^2\frac{m_2}{m_3} 
\hspace{14mm}
\nn \\
+ \left( \frac{1}{m_2^2}\!-\!\frac{1}{2m_3^2}
                    \!-\!\frac{1}{2m_1^2} \right) 
         \ln^2\frac{m_1}{m_3}
+ \frac{4\pi^2}{3} \left( \frac{1}{m_1^2} + \frac{1}{m_2^2}
                          + \frac{1}{m_3^2} \right)
\hspace{23mm}
\nn \\
\left. \left. 
- 4\pi \left( \frac{\theta_1}{m_1^2} 
                          \!+\! \frac{\theta_2}{m_2^2}
                          \!+\! \frac{\theta_3}{m_3^2} \right)
+ 4\pi \left(\frac{1}{m_1}\!+\!\frac{1}{m_2}\!+\!\frac{1}{m_3} 
               \right)
  \sqrt{\frac{m_1\!+\!m_2\!+\!m_3}{m_1 m_2 m_3}}
\right]
\right\} \!+\! {\cal{O}}(\ep) 
\eea

The presented expressions explicitly obey all required symmetries. 
In particular, the result (\ref{L4112t}) is symmetric under
$m_1\leftrightarrow m_2$, whereas eqs.~(\ref{L6222t}) and
(\ref{L4111t}) are totally symmetric in $m_1,m_2,m_3$.

%===========================================================
\section{Some special cases}
%===========================================================

When two of the masses are equal (say, $m_1=m_2\equiv m$, 
$m_3\equiv M$) 
the presented general expressions can be simplified.
Let us consider, as an example, the integral $L(4-2\ep;1,1,1)$
and introduce a dimensionless variable $\mu \equiv M/m$.

At two different pseudothresholds,  $k^2=(2m-M)^2=(2-\mu)^2 m^2$ and
$k^2=M^2=\mu^2 m^2$, we get the following results:
\bea
\label{pseudo1-111}
\left. L(4-2\ep;1,1,1)
\frac{}{}\right|_{\begin{array}{c}
                  {}_{m_1=m_2\equiv m, \;\; m_3\equiv M = \mu m}\\
                  {}^{k^2=(2m-M)^2=(2-\mu)^2 m^2}
                  \end{array}}
= \pi^{4-2\ep}\; \frac{\Gamma^2(1+\ep)}{(1-\ep)(1-2\ep)} \; 
m^{2-4\ep}
\hspace{22mm}
\nn \\
\times \! \left\{
- \frac{2+\mu^2}{2\ep^2} + \frac{(2-\mu)^2}{4\ep}
+ \frac{2\mu^2}{\ep} \ln\mu + \frac{1}{2}(2+\mu^2)
+ \frac{7}{8} (2-\mu)^2
+ \frac{6-2\mu+\mu^2}{2-\mu} \mu \ln\mu
\right.
\nn \\
\left.
-2\mu^2 \ln^2\mu
-\frac{4(1\!-\!\mu)^2 (1\!+\!\mu)(3\!-\!\mu)}{(2-\mu)^2}
\left[ \Li{2}{1\!-\!\mu} + \frac{\pi^2}{12} \right]
\right\} + {\cal{O}}(\ep),
\eea
\bea   
\label{pseudo2-111}
\left. L(4-2\ep;1,1,1)
\frac{}{}\right|_{\begin{array}{c}
                  {}_{m_1=m_2\equiv m, \;\; m_3\equiv M = \mu m}\\
                  {}^{k^2=M^2=\mu^2 m^2}
                  \end{array}}
= \pi^{4-2\ep}\; \frac{\Gamma^2(1+\ep)}{(1-\ep)(1-2\ep)} \; 
m^{2-4\ep}
\hspace{22mm}
\nn \\
\times \left\{
- \frac{2+\mu^2}{2\ep^2}
+ \frac{\mu^2}{4\ep}
+ \frac{2}{\ep} \mu^2 \ln\mu +1 + \frac{11}{8} \mu^2
- (2+\mu^2) \ln\mu - 2 \mu^2 \ln^2\mu
\right.
\nn \\ 
\left.
+ \frac{(1-\mu^2)^2}{\mu^2}
\left[ \Li{2}{1-\mu^2} - \frac{\pi^2}{6} \right]
\right\} + {\cal{O}}(\ep) .
\eea
Note that the latter expression is relevant for the on-shell 
calculations
and corresponds to the results presented in refs.~\cite{BG,CzM}.

At the threshold, when $k^2=(2m+M)^2=(2+\mu)^2 m^2$, we get
\bea
\label{thr-111}
\left. L(4-2\ep;1,1,1)
\frac{}{}\right|_{\begin{array}{c}
                  {}_{m_1=m_2\equiv m, \;\; m_3\equiv M = \mu m}\\
                  {}^{k^2=(2m+M)^2=(2+\mu)^2 m^2}
                  \end{array}}
= \pi^{4-2\ep}\; 
\frac{\Gamma^2(1+\ep)}{(1-\ep)(1-2\ep)} \; m^{2-4\ep}
\hspace{27mm}
\nn \\ 
\times \! \left\{
- \frac{2+\mu^2}{2\ep^2} + \frac{(2+\mu)^2}{4\ep}
+ \frac{2\mu^2}{\ep} \ln\mu + \frac{1}{2}(2+\mu^2)
+ \frac{7}{8} (2+\mu)^2
\right.
\hspace{44mm}
\nn \\
- \frac{6+2\mu+\mu^2}{2+\mu} \mu \ln\mu
-2\mu^2 \ln^2\mu
- \frac{4\mu^{1/2}}{(2+\mu)^{3/2}}(3+2\mu+\mu^2) \pi
\hspace{45mm}
\nn \\
\left.
+\frac{4(1\!+\!\mu)^2 (1\!-\!\mu)(3\!+\!\mu)}{(2+\mu)^2}\!
\left[ \Li{2}{\frac{1}{1\!+\!\mu}}
\!+\!\frac{1}{2} \ln^2(1\!+\!\mu)
\!+\! \pi \arccos\left(\!\frac{1}{1\!+\!\mu}\!\right)
\!-\! \frac{5\pi^2}{12} \right]   
\right\} \!+\! {\cal{O}}(\ep) .
\eea
                  
For $\mu=1$, i.e. when all masses are equal, the results
(\ref{pseudo1-111}) and (\ref{pseudo2-111})
give the same as eq.~(30) of the paper \cite{BFT} (see also in
\cite{Broadh}),
\be               
\left. L(4-2\ep;1,1,1)
\right|_{\begin{array}{c} {}_{m_1=m_2=m_3=m} \\
                          {}^{k^2=m^2} \end{array} }
= \pi^{4-2\ep}\frac{\Gamma^2(1+\ep)}{(1-\ep)(1-2\ep)} m^{2-4\ep}
\left\{ -\frac{3}{2\ep^2}+\frac{1}{4\ep}+\frac{19}{8}
\right\} + {\cal{O}}(\ep),
\ee
whereas eq.~(\ref{thr-111}) yields the threshold value of
the sunset integral with equal masses 
\be
\label{mmm_thr}
\left. L(4\!-\!2\ep;1,1,1)
\right|_{\begin{array}{c} \!{}_{m_1=m_2=m_3=m}\! \\
                          {}^{k^2=9m^2} \end{array} }
\!\! = \pi^{4-2\ep}\frac{\Gamma^2(1+\ep)}{(1\!-\!\ep)(1\!-\!2\ep)}
m^{2-4\ep}
\left\{ -\frac{3}{2\ep^2}\!+\!\frac{9}{4\ep}\!+\!\frac{75}{8}
\!-\!\frac{8\pi}{\sqrt{3}}
\right\}          
+ {\cal{O}}(\ep).
\ee
We have also reproduced the same result (\ref{mmm_thr})
by using the dispersion technique.

%===========================================================
\section{Conclusions}
%===========================================================

We have obtained analytic results for the threshold and 
pseudothreshold values of the sunset diagram with arbitrary masses.
The results are expressed in terms of dilogarithms of ratios
of the masses, all other functions being elementary. The results
are explicitly
symmetric with respect to all symmetries they have to obey.
Certain checks, including numerical ones, confirm that the
results are self-consistent.
Note that most of the terms in eqs.~(\ref{L6222t})--(\ref{L4111t})  
look just as naive analytic continuation of the results
(\ref{L6222p})--(\ref{L4111p}), if one changes
$m_3$ into $(-m_3)$. 
The non-trivial terms are those involving $\pi$, including the
$\theta_i$ terms.

We note that the complete (unsubtracted) results, involving
the $1/\ep^2$ and $1/\ep$ ultraviolet poles, are simpler
than the finite function obtained by subtraction of two
terms of the expansion in $k^2$ near $k^2=0$ (cf. e.g. in \cite{PT}).
In the latter case, we would also get dilogarithms (or
Clausen functions) corresponding to the vacuum diagrams
\cite{vacuum,DT1}. In particular, for the equal-mass case
(\ref{mmm_thr}) the $\mbox{Cl}_2(\pi/3)$ contribution is missing in 
eq.~(\ref{mmm_thr}). 
 
As an extension of the presented approach, two-loop self-energy 
diagrams with four and five propagators may also be considered.
For example, the exact result for the pseudothreshold
value of the ``master'' diagram with equal masses was
presented in \cite{BJ}. Although the results obtained may be used 
for all purposes listed in the introduction, we would like to 
emphasize that their application to constructing expansions near 
the thresholds is one of the most important points.

\vspace{3mm}

{\bf  Acknowledgements.}
We would like to thank A.G.~Grozin, V.A.~Smirnov and J.B.~Tausk 
for useful discussions and help in checking results 
for some special cases. 
A.~D. and N.~U. are grateful to the Instituut-Lorentz,
University of Leiden for their hospitality during our visits.
The research was mainly supported by the EU grant INTAS-93-0744.
A partial support from the RFBR grants 95-02-05794 (N.~U.) 
and 96-01-00654 (A.~D.) is acknowledged.


\begin{thebibliography}{99}
 
\bibitem{dimreg} G.~'t~Hooft and M.~Veltman,
    {\em Nucl.Phys.} B44 (1972) 189;\\
    C.G.~Bollini and J.J.~Giambiagi, {\em Nuovo Cim.} 12B (1972) 20.

\bibitem{Tar_new} 
G.~Weiglein, R.~Scharf and M.~B\"{o}hm,
{\em Nucl.~Phys.} B416 (1994) 606; \\
O.V.~Tarasov, {\em Nucl.~Phys.} B502 (1997) 455; \\
A.~Ghinculov and Y.-P.~Yao, Freiburg preprint THEP-97-04
             (hep-ph/9702266). 

\bibitem{special} 
G.~K\"all\'en and A.~Sabry,
{\em Dan.~Mat.~Fys.~Medd.} 29, No.17 (1955) 1; \\  
D.J.~Broadhurst, {\em Phys. Lett.} B101 (1981) 423; \\
T.H.~Chang, K.J.F.~Gaemers and W.L.~van Neerven, {\em Nucl. Phys.}
  B202 (1982) 407; \\
A.~Djouadi, {\em Nuovo Cim.} 100A (1988) 357; \\
B.A.~Kniehl, {\em Nucl. Phys.} B347 (1990) 86.
A.V.~Kotikov, {\em Phys.~Lett.} B254 (1991) 158;
{\em Mod. Phys. Lett.} A6 (1991) 677; \\
P.N.~Maher, L.~Durand and K.~Riesselmann,
{\em Phys.~Rev.} D48 (1993) 1061; \\
D.T.~Gegelia, K.Sh.~Japaridze and K.Sh.~Turashvili,
{\em Teor. Mat. Fiz.} 101 (1994) 225.

\bibitem{Broadh} D.J.~Broadhurst,
{\em Z.~Phys.} C47 (1990) 115.   

\bibitem{BFT}
D.J.~Broadhurst, J.~Fleischer and O.V.~Tarasov,
{\em Z.~Phys.} C60 (1993) 287.

\bibitem{Scharf}
R.~Scharf, Diploma thesis (W\"urzburg, 1991); Doctoral thesis
(W\"urzburg, 1994); \\
R.~Scharf and J.B.~Tausk, {\em Nucl.~Phys.} B412 (1994) 523.

\bibitem{Rajantie} A.K.~Rajantie,
{\em Nucl.~Phys.} B480 (1996) 729.

\bibitem{BDS}
F.A.~Berends, A.I.~Davydychev and V.A.~Smirnov,
{\em Nucl.~Phys.} B478 (1996) 59.

\bibitem{Tkachov} F.V.~Tkachov, e-print hep-ph/9703424. 

\bibitem{BS} M.~Beneke and V.A.~Smirnov,
Preprint CERN-TH-97-315 (hep-ph/9711391)

\bibitem{BG} D.J.~Broadhurst and A.G.~Grozin,
{\em Phys.~Rev.} D52 (1995) 4082.

\bibitem{CzM} A.~Czarnecki and K.~Melnikov, Karlsruhe
preprint TTP-97-08 (hep-ph/9703277).

\bibitem{GvdB} A.~Ghinculov and J.J.~van der Bij,
{\em Nucl.~Phys.} B436 (1995) 30.

\bibitem{numeric}
D.~Kreimer, {\em Phys.~Lett.} B273 (1991) 277; \\
J.~Fujimoto, Y.~Shimizu, K.~Kato and Y.~Oyanagi, 
    KEK preprint 92-213;\\
F.A.~Berends and J.B.~Tausk, {\em Nucl.~Phys.} B421 (1994) 456;\\
F.A.~Lunev, {\em Phys.~Rev.} D50 (1994) 7735; \\
A.~Czarnecki, U.~Kilian and D.~Kreimer,
   {\em Nucl.~Phys.} B433 (1995) 259; \\
S.~Bauberger, F.A.~Berends, M.~B\"{o}hm and M.~Buza,
   {\em Nucl.Phys.} B434 (1995) 383; \\
S.~Bauberger and M.~B\"{o}hm, {\em Nucl.~Phys.} B445 (1995) 25.

\bibitem{PT} P.~Post and J.B.~Tausk,   
{\em Mod.~Phys.~Lett.} A11 (1996) 2115.

\bibitem{DT1} A.I.~Davydychev and J.B.~Tausk,
{\em Nucl.~Phys.} B397 (1993) 123.

\bibitem{FT} J.~Fleischer and O.~V.~Tarasov,
   {\em Z.~Phys.} C64 (1994) 413.

\bibitem{Tar} O.V.~Tarasov, {\em Nucl.~Phys.} B480 (1996) 397. 

\bibitem{DST} A.I.~Davydychev, V.A.~Smirnov and J.B.~Tausk,
{\em Nucl.~Phys.} B410 (1993) 325.

\bibitem{BBBS}
F.A.~Berends, M.~Buza, M.~B\"{o}hm and   
R.~Scharf, {\em Z.~Phys.} C63 (1994) 227.

\bibitem{Mendels} E.~Mendels, {\em Nuovo~Cim.} 45A (1978) 87.

\bibitem{DT2}
A.I.~Davydychev and J.B.~Tausk, {\em Phys.~Rev.} D53 (1996) 7381.

\bibitem{ChengWu} H.~Cheng and T.T.~Wu, {\em Expanding
protons: scattering at high energies} (MIT press, Cambridge,
Massachusetts, 1987).

\bibitem{BFO2} K.S.~Bj{\o}rkevoll, G.~F\"aldt and P.~Osland,
{\em Nucl.~Phys.} B386 (1992) 303.

\bibitem{DD} A.I.~Davydychev and R.~Delbourgo,
Preprint UTAS-PHYS--97-12 (hep-th/9709216).

\bibitem{vacuum}
J.J.~van der Bij and M.~Veltman, 
  {\em Nucl.~Phys.} B231 (1984) 205; \\
F.~Hoogeveen, {\em Nucl.~Phys.} B259 (1985) 19; \\
J.J. van der Bij and F.~Hoogeveen, 
  {\em Nucl.~Phys.} B283 (1987) 477; \\
C.~Ford, I.~Jack and D.R.T.~Jones,
{\em Nucl.Phys.} B387 (1992) 373.

\bibitem{BJ}
V.~Borodulin and G.~Jikia, {\em Phys.~Lett.} B391 (1997) 434.

\end{thebibliography}
\end{document}